\documentclass[a4paper,aps,prl,twocolumn,preprintnumbers,amsmath,amssymb]{revtex4}
\usepackage{amsmath}
\usepackage{verbatim}
\usepackage{graphicx}

\usepackage{color}

 \newcommand{\ket}[1]{\left|#1\right\rangle} 
 \newcommand{\bra}[1]{\left\langle#1\right|} 

\begin{document}
 
\title{Hybrid annealing using a quantum simulator coupled to a classical computer}

\author{Tobias Gra{\ss}$^{1,2}$ and Maciej Lewenstein$^{2,3}$}
\affiliation{$^1$Joint Quantum Institute, University of Maryland, College Park, MD 20742, U.S.A.}
\affiliation{$^2$ICFO-Institut de Ci\`encies Fot\`oniques, The Barcelona Institute of Science and Technology, 08860 Castelldefels, Spain}
\affiliation{$^3$ICREA-Instituci\'o Catalana de Recerca i Estudis Avan\c{c}ats, Llu\'is Companys 23, 08010 Barcelona, Spain}

\begin{abstract}
Finding the global minimum in a rugged potential landscape is a computationally hard task, often equivalent to relevant optimization problems. Simulated annealing is a computational technique which explores the configuration space by mimicking thermal noise. By slow cooling, it freezes the system in a low-energy configuration, but the algorithm often gets stuck in local minima. In quantum annealing, the thermal noise is replaced by controllable quantum fluctuations, and the technique can be implemented in modern quantum simulators. However, quantum-adiabatic schemes become prohibitively slow in the presence of quasidegeneracies. Here we propose a strategy which combines ideas from simulated annealing and quantum annealing. In such hybrid algorithm, the outcome of a quantum simulator is processed on a classical device. While the quantum simulator explores the configuration space by repeatedly applying quantum fluctuations and performing projective measurements, the classical computer evaluates each configuration and enforces a lowering of the energy. We have simulated this algorithm for small instances of the random energy model, showing that it potentially outperforms both simulated thermal annealing and adiabatic quantum annealing. It becomes most efficient for problems involving many quasi-degenerate ground states. 
\end{abstract}

\maketitle


\begin{figure}[h]
\centering
\includegraphics[width=0.48\textwidth, angle=0]{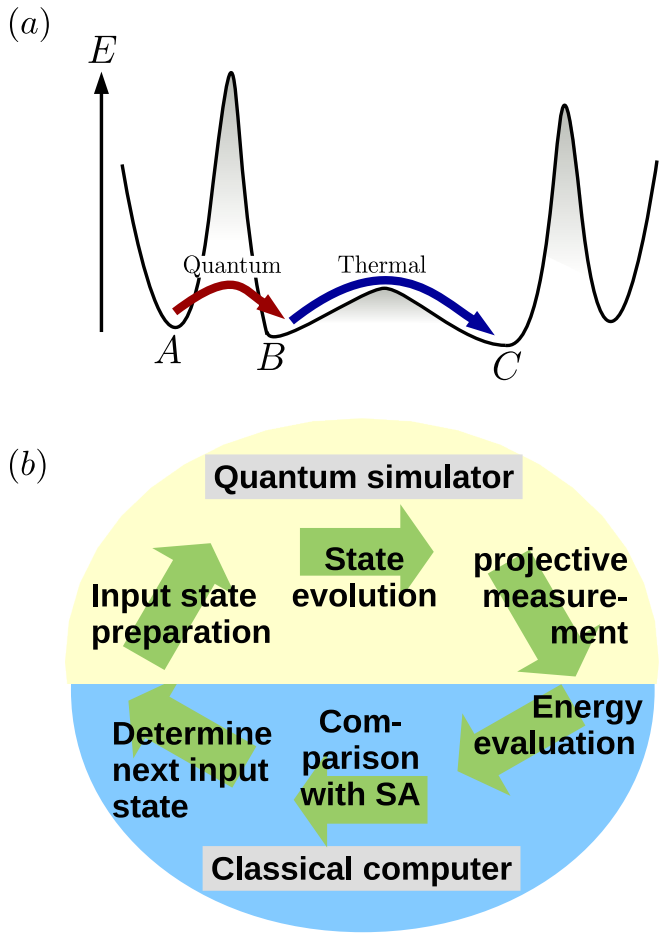}
\caption{\label{schematic} 
(a) In the depicted energy landscape, the global minimum at $C$ can be reached from $A$ via a combination of thermal annealing and quantum tunneling. 
The high potential barrier from $A$ to $B$ can be tunneled quantum-mechanically, and the wide barrier between $B$ and $C$ can be surpassed by thermal fluctuations. 
(b) One annealing cycle in the hybrid algorithm: given an input state from the computational basis, a quantum simulator searches for nearby minima by time-evolving the state in the presence of quantum fluctuations. After a projective measurement, a classical device evaluates the energy of the output state. Based on this evaluation, which might also take into account the result of a simultaneously performed simulated annealing (SA) scheme, a new input state for the next cycle is defined.}
\end{figure}

\textit{Introduction.}
Constructing a quantum computer has been an outstanding goal and immense driving force for physical research during the past decades. In recent years, a series of notable advances has been made towards this goal: Using superconducting qubits several companies, such as D-Wave \cite{dwave11}, IBM \cite{ibm12,ibm_web}, and Google \cite{vasco}, have developed information-processing devices. Academic research has produced prototypes of ion-based quantum computers \cite{lanyon11,debnath16}. Other systems which may allow for performing quantum computations are neutral atoms, such as Rydberg atoms \cite{isenhower10}, atoms in cavities \cite{ritter12} or photonic crystals \cite{douglas15}, as well as quantum dots or artificial atoms implemented as defects in a solid-state system \cite{gao15,yao12}. Apart from finding the best amongst these different hardware architectures, research also has to determine the algorithmic design of future quantum computers. Universal quantum computers, as realized in Refs. \cite{lanyon11,debnath16,vasco} or even accessible via a cloud service \cite{ibm_web}, break any computational task into simple gate operations. While this is a powerful approach, in particular as it also admits quantum error correcting codes \cite{devitt13}, the best scalability so far has been achieved with so-called analog quantum computers \cite{dwave11}. These are specific-purpose devices, designed to address one problem of interest. As there are mappings between different problems, such special-purpose computer would still be useful for addressing a whole class of problems. For instance, if an analog quantum computer can solve just a single NP-complete problem in polynomial time, by the definition of NP-completeness any NP problem would become solvable in polynomial time. Such quantum device would then outperform any known classical algorithm for a diverse variety of problems, containing the traveling salesman problem, various satisfiability problems, graph-coloring problems, the number partitioning problem, to name just a few. 

Typically, analog quantum computers implement an algorithm known as quantum annealing \cite{kadowaki98} or adiabatic quantum computing \cite{farhi01}. The idea behind this is the following: Starting from an initial Hamiltonian with an easy-to-prepare ground state one can reach the ground state of a complex target Hamiltonian by slowly varying a Hamiltonian parameter. The target Hamiltonian should be chosen such that its ground state provides the solution to the computational problem of interest. Here, one can exploit the NP-completeness of Ising spin glass models \cite{barahona,lucas}, allowing for any NP problem to be mapped onto an Ising model. In this Ising picture, the complicated target Hamiltonian is usually approached from an initial Hamiltonian, in which a strong magnetic field simply polarizes all spins in a transverse direction.

However, the computational time needed for the adiabatic computation increases quadratically with the inverse energy gap. For several relevant models, the gap has been shown to become exponentially small on the annealing path \cite{joerg08,altshuler10}. In these cases, approaches which do not rely on the adiabatic principle are desirable. In this paper, we describe an annealing scheme which avoids the requirement of adiabaticity. Instead, we envisage a hybrid solution where classical information processing is paired with quantum technologies, see Fig.~\ref{schematic}(b). More concretely, the quantum part of our protocol consists of (i) the preparation of an input state chosen from the computational basis, (ii) a short unitary time evolution of this state, performed by a quantum simulator which implements the target Hamiltonian exposed to quantum fluctuations, (iii) a measurement which projects the evolved quantum state onto a state from the computational basis. The outcome of this measurement is then processed to a classical computer which evaluates the configuration, and determines the next input state. In the simplest version, this evaluation may be restricted to comparing the energy in the initial and the final configuration. Repeating these steps allows for systematically lowering the energy of the system.

Such protocol resembles a simulated annealing algorithm which was pioneered in Ref. \cite{kirkpatrick83}. Simulated annealing mimics the evolution of a classical system in the presence of thermal noise, for instance implemented by random spin flips. Updates are accepted or rejected according to a probability distribution characterized by some temperature. Slowly decreasing the temperature maximizes the probability of reaching a low-energy state. In the protocol described above, thermal noise is replaced by quantum fluctuations. As illustrated in Fig.~\ref{schematic}(a), such an algorithm based on quantum fluctuations might surpass high energy barriers for which the likelihood of escaping is strongly suppressed in the case of thermal annealing. In contrast, surpassing wide energy barriers might be no problem for thermal noise, and combining both quantum and thermal fluctuations would enhance the chances of finding the global minimum. In the scheme described above, such combination can straightforwardly be achieved by running a classical simulated annealing algorithm in parallel to the quantum simulation.

As a hardware requirement to run such algorithm one needs a quantum simulator which implements the Hamiltonian of interest (e.g. a classical Ising-like model), and exposes it to quantum fluctuations (e.g. to a transverse magnetic field). Moreover, one should be able to initialize the quantum simulator in an arbitrary classical state, and to project the evolved quantum state back onto a state from the classical basis via a projective measurement. All information processing is done by a classical computer, that is, evaluating the energies (corresponding to the cost function of the optimization problem), deciding about whether to reject or accept the measured state, etc. The mentioned requirements to the quantum simulator are already met, for instance, in existing trapped ion experiments \cite{richerme14,jurcevic14}. There also exist proposals how to implement NP-complete problems in an ionic quantum simulator \cite{hauke15,grass16}.

In this manuscript, we numerically test the envisaged protocol. To do so, we focus on a random energy model, for which simulated annealing fails because similar spin configurations may have completely different energies. In such random energy landscape, also adiabatic quantum annealing has been shown to suffer from an exponentially small gap \cite{joerg08}. After introducing the model, and describing the details of our algorithm, we simulate the algorithm in small Hilbert spaces of dimension up to $2^{12}$, and compare it to other annealing strategies. For the random energy model, our annealing scheme, even if it is restricted to quantum fluctuations only, outperforms the simulated thermal annealing. It can further be improved by taking into account also thermal fluctuations. However, for the small-sized problems considered here, the quantum adiabatic annealing turns out to be the best choice. The situation changes if we artificially introduce quasi-degeneracies in the energy landscape. While this even enhances the performance of our hybrid algorithm, it prevents the adiabatic scheme from finding the correct state. In a final discussion, we point out the main advantages of the algorithm.

\textit{Model.}
In the random energy model a $D$-dimensional Hilbert space, spanned by states $\ket{\alpha}$, is associated with a random energy landscape, $H_0=\sum_\alpha E_\alpha \ket{\alpha}\bra{\alpha}$, where  the $E_\alpha$ are chosen from some probability distribution. Here, we consider the Hilbert space of $N$ spin-1/2 particles, $D=2^N$, and the energies are normally distributed with mean zero, $\langle E_\alpha \rangle =0$, and variance $\langle E_\alpha^2 \rangle = \epsilon^2 = 1$, introducing a unit $\epsilon$ for energy. Under these premises, the random energy model provides a simple toy model for a spin glass transition. Above a critical temperature, an exponentially large number of states contributes almost equally to the free energy, while in a low-temperature phase the free energy is dominated by the ground state \cite{derrida80}.

Choosing the $\ket{\alpha}$ to be eigenstates of $\sigma_z^{(i)}$ operators, we introduce quantum fluctuations through a transverse field term $H_1=B\sum_i \sigma_x^{(i)}$. While in an adiabatic annealing algorithm, the field strength $B$ decreases with time, starting from a large value which polarizes all spins along $\sigma_x$, our hybrid algorithm will evolve an initial state $\ket{\alpha}$ under a constant Hamiltonian $H=H_0+H_1$.  The field strength $B$ is supposed to be of the order of $\epsilon$. On the one hand, such field is sufficiently weak, such that it does not disturb very much the classical energy landscape for which we are interested to find its minimum. On the other hand, it is strong enough to introduce enough quantum fluctuations for a fast tunneling to other classical states. Accordingly, the proper choice enhances our chance of finding a new local minimum after projecting the time-evolved state onto the classical basis. A second parameter which is crucial for the performance of our algorithm is the duration of the time evolution. On  short time scales, the probability of reaching a configuration of less energy increases with the evolution time $t$. After some time, however, the improvement rate oscillates around an average value, and, on average, one does not benefit from a longer evolution anymore. In our simulations, a good timing for making the projective measurement was found to be around $t=10$ (in units $\tau \equiv \hbar/\epsilon$).

In a practical application, the classical energy landscape shall reproduce the cost function of a given optimization problem. In this case, evaluation of the energies can easily be done using a classical computer. If the quantum simulator, after the projective measurement, produces an output with decreased energy, the algorithm will accept this output as the new input state. Otherwise, it shall restore the previous configuration, and repeat the time evolution under the quantum field. In the latter case, even if the parameters $B$ and $t$ remain unchanged, the subsequent output may be different from the previous one, since the projective measurement introduces intrinsic randomness to the algorithm. Nevertheless, by varying one of these parameters, the performance of the algorithm is strongly enhanced.  In our simulation, we therefore have chosen a random field $B$ in the interval $[0,1]$. As in simulated annealing, the algorithm also benefits from a non-zero acceptance rate for steps towards higher energies, to avoid getting stuck in metastable configurations. In our simulation, we have chosen an acceptance probability $p(\Delta E)=\min\{\exp(-\beta \Delta E),1\}$ with $\Delta E$ the energy difference between new and old configuration. We worked with a constant low temperature, $\beta=10/\epsilon$.

As described so far, our algorithm only involves quantum fluctuations, but it resembles a simulated annealing scheme in how information is processed. We therefore refer to it as ``hybrid annealing'' (HA), to distinguish it from the traditional ``simulating annealing'' (SA) where updates are produced by mimicking thermal noise. An algorithm where hybrid annealing and simulating annealing are run in parallel will be called ``hybrid-simulating annealing'' (HA+SA). We first consider the performance of the HA algorithm, and then compare it with other strategies.

\textit{Results.}
In our simulation of the HA algorithm, we need to mimic the time evolution of a quantum system on a classical computer. This necessarily restricts our simulation to small samples, and we have considered Hilbert spaces spanned by $N=8$ up to $N=12$ qubits. Despite the self-averaging nature of the random energy model, we therefore need to consider averages over many instances. The average over many instances will also compensate for the random nature of the algorithm itself, due to the fact that it consists of repeated projective measurements.

First, we have determined how many evolution steps it takes until the ground state is reached. The results are presented in Table \ref{table1}. Although the annealing time scales exponentially with $N$, it grows significantly slower than the size of the Hilbert space: Both average and median number of steps fit well to $f(N) \propto b^{N-8}$, with $b=1.60 \pm 0.02$. In contrast, a random search would require an average number of steps that grows linearly with the Hilbert space, that is, $\propto 2^N$.
\begin{table}[t]
\vspace{0.5cm}
\begin{tabular}{c|c|c}
$N$ & average number of steps & median number of steps \\
\hline
8 & 100 & 60 \\
9 & 160 & 100 \\
10 & 240 & 150 \\
11 & 390 & 250 \\
12 & 650 & 390
\end{tabular}
\caption{\label{table1} For different Hilbert space sizes (parametrized by the number of qubits $N$), we list the average and median number of annealing steps needed by the hybrid algorithm to reach the ground state. For the averaging, we have taken into account 1000 instances of the random energy model (667 in the case of $N=12$). In each run, the algorithm was initialized in the middle of the energy spectrum, i.e. in the $2^{N-1}$th level.
}
\end{table}

Next, we compare the performance of the HA algorithm to other annealing approaches. To this aim, our simulation now fixes the runtime, and we compare the different algorithms by evaluating their success rate. In algorithms consisting of repeated updating steps (HA, SA, HA+AS), we fix the number of steps to 200. For the HA, as detailed above, this amounts to a minimum total runtime of $t_1=2000\tau$, where instantaneous state preparation and immediate classical data processing are assumed. We can then directly compare the HA runtime to the runtime of a quantum adiabatic annealing approach. In the latter, a single, uninterrupted quantum time evolution is performed for some given amount of time. Its final outcome is the evolved quantum state at time $t_1$, and the success probability is given by the squared overlap of this state with the ground state of $H_0$. 

Apart from fixing the runtime, other details determine the protocol for quantum adiabatic annealing: The standard strategy is to prepare the system in an initial state which is fully polarized by a transverse field term $H_1$, and evolve under a time-dependent Hamiltonian $H(t)= \alpha_0(t) H_0 + \alpha_1(t) H_1$. In our simulation, we have considered a constant target Hamiltonian, i.e. $\alpha_0(t)={\rm const.}=1$, and an exponentially decaying quantum field $\alpha_1(t)=\exp(-t/t_0)$, with $t_0=t_1/10$. The field strength in $H_1$ is $B=10\epsilon$. The success probability $p_{\rm AA}$ is given by the squared overlap of the final state with the desired target state. 
Averaged over 100 realizations with $N=10$ qubits, we find $p_{\rm AA}=0.83$. This is better than the corresponding success probability in the hybrid scheme, $p_{\rm HA}=0.71$ (averaged over 1000 runs), obtained by counting the number of times the ground state is reached within the given runtime.

However, it is known that the adiabatic annealing becomes prohibitively expensive for large $N$ due to exponentially small gaps \cite{joerg08}. In an attempt to mimic this problem in a small sample, we have artificially introduced a quasi-degenerate manifold above the true ground state. This makes it impossible for an adiabatic algorithm to determine the true ground state. In our simulation, we have chosen the four lowest excitations to have energies $E=E_{\rm GS} + 0.001\epsilon$. This reduces the success rate of the adiabatic scheme to $p_{\rm AA}=0.26$. On the other hand, the performance of the hybrid annealing scheme is even improved, $p_{\rm HA}=0.88$. 

\begin{figure}
\centering
\includegraphics[width=0.48\textwidth, angle=0]{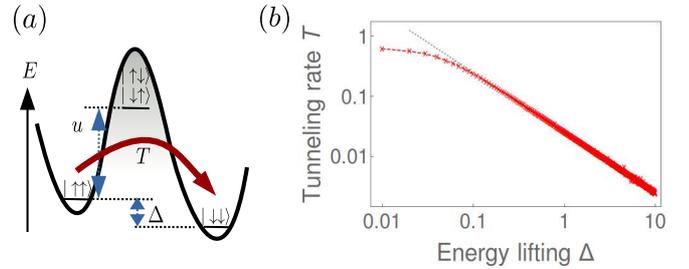}
\caption{\label{tunnel} 
(a) An energy landscape for two spins: The global minimum $\ket{\downarrow\downarrow}$ is detuned from the local minimum $\ket{\uparrow\uparrow}$ by an energy offset $\Delta$, and separated by a barrier of height $u$. (b) For the depicted energy landscape (with $u=100$), we plot the tunneling probability $T$ in a transverse field of strenth $B=1$ as a function of the detuning $\Delta$. Asymptotically, $T\sim1/\Delta$, as indicated by the fit (dashed line). The model explains the enhanced performance of our algorithm in quasi-degenerate energy landscapes.}
\end{figure}

This improvement can be understood from an enhanced tunneling between levels at equal energies, and provides a major benefit of the hybrid algorithm compared to adiabatic annealing schemes. The behavior is captured by a simple two-spin model illustrated in Fig.~\ref{tunnel}(a): In this model, the local minimum $\ket{\uparrow \uparrow}$ is separated from the global minimum $\ket{\downarrow\downarrow}$ by a high energy cost $u$ in configurations $\ket{\uparrow\downarrow}$ and $\ket{\downarrow\uparrow}$. Within a transverse field $B$, there is a finite probability of tunneling from the local to the global minimum, $p(t)=|\bra{\downarrow\downarrow}e^{iHt}\ket{\uparrow\uparrow}|$. In Fig.~\ref{tunnel}(b), we plot the time-averaged tunneling rate $T=\lim_{\tau\rightarrow \infty} \frac{1}{\tau}\int_{0}^{\tau} p(t){\rm d}t$ as a function of the detuning $\Delta$. Asymptotically, the tunneling rate $T$ is inversely proportional to the energy offset $\Delta$. This observation explains the enhanced performance of our algorithm in a quasi-degenerate scenario.

\begin{figure}[t]
\centering
\includegraphics[width=0.48\textwidth, angle=0]{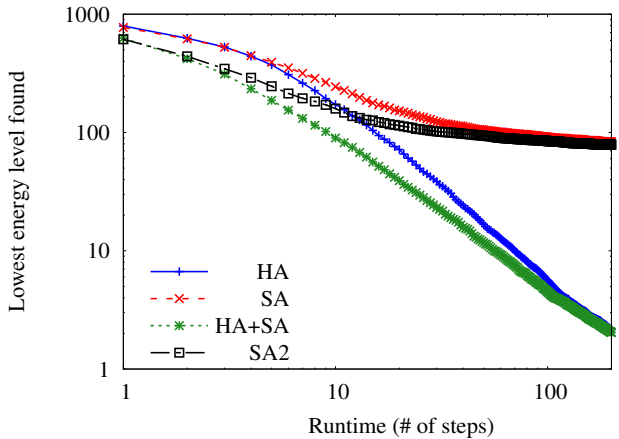}
\caption{\label{lev} 
 Performance of different algorithms applied to the random energy model with $2^{11}=2048$ levels: Averaged over 1000 instances, we plot the lowest level which each algorithm has found as a function of the runtime, that is, of the number of update steps the algorithm has gone through. Each algorithm starts from the 1024th state. At any runtime, the combination of hybrid annealing and simulated annealing (HA+SA) gives the best result. On the long run, the hybrid algorithm (HA) performs equally well, whereas the two purely thermal algorithms, SA and SA2, remain stuck in relatively high levels.
}
\end{figure}

We have compared the hybrid scheme also with the outcome of simulated annealing (SA), where updates are produced by randomly flipping one spin. As HA, the SA scheme accepts an update following a probability distribution $p(\Delta E,\beta)=\min\{\exp(-\beta \Delta E),1\}$.  However, to bring the system from one potential well to another one, in the SA scheme one typically needs a series of accepted updates (of the order $N$ flips). To make such a process likely, a high temperature is required in the beginning. In the course of the evolution, the temperature is then slowly decreased in order to relax into a  (meta)stable configuration. Thus, in contrast to the HA algorithm, for SA it is crucial that the parameter $\beta$ is varied during the algorithm. In the data shown in Fig.~\ref{lev}, we use $\beta(k)=\beta_0 c^{-k}$, with $\beta_0=1/\epsilon$ and $c=0.98$. Starting from $k=0$, the variable $k$ increases in steps of one after ten successful updates, or after twenty steps. We find that in only less than ten percent of the instances, the SA scheme is able to find the ground state of the random energy model within 200 steps. In most cases, it quickly gets stuck in local minima for long times. Clearly, one reason for the bad performance of the simulated annealing is the lack of any structure in the random energy landscape.

An advantage of the HA scheme is the fact that it can easily be combined with purely classical algorithms. For instance,  
we are able to improve the HA algorithm by combining it with the SA algorithm. As depicted in Fig.~\ref{schematic}(a), high barriers between
two minima suppress thermal tunneling and make simulated annealing fail. On the other hand, large energy offsets between minima can reduce quantum tunneling. Following this picture, an algorithm which allows for both processes should have the best performance. 

Accordingly, we have designed an algorithm which performs a hybrid-simulated annealing protocol (HA+SA). In each step, this algorithm chooses between two updates: One update is based on a random spin flip mimicking thermal fluctuations. A second update is produced by the unitary time evolution in a quantum field and the projective measurement in the hybrid algorithm. Among these two updates, the HA+SA algorithm chooses the one of lower energy, which is then accepted or not depending on the energy balance.

In Fig. \ref{lev} we  analyze the performance of each algorithm. As a function of the number of steps, the figure plots the index of the lowest energy level reached so far. From this figure, it is seen that simulated annealing (SA) gets stuck in a relatively high level (on average 83rd level after 200 steps for $N=11$), while the hybrid annealing (HA) on average reaches the first-excited level. The combination of hybrid annealing and simulated annealing (HA+SA) yields only a slightly better performance on long time scales, but its advantages are apparent on short time scales. Here, HA+SA is the scheme for which the energy drops fastest. Since the combined algorithm produces two updates per step, we may compare it with a simulated annealing algorithm which chooses the better update among two random spin flips (SA2). It is seen that SA2 improves the simulated annealing performance on short time scales, but it is inefficient on the long run.

\textit{Discussion.}
Summarizing the results presented above, our simulations show that the proposed hybrid annealing scheme outperforms the traditional simulated annealing approach on the random energy model, though without achieving polynomial scaling. The algorithm can be combined with the simulated annealing approach, giving rise to a setup where classical and quantum algorithm supplement each other (HA+SA). Accordingly, the HA+SA scheme will always outperform a strategy based on SA alone. 

Comparing the hybrid approach with a quantum-adiabatic scheme, the latter shows better success rates in a small sample of the random energy model. However, small samples favor the adiabatic scheme due to pronounced finite-size gaps. Therefore, we have artificially increased the complexity by introducing quasi-degeneracies to the otherwise randomly generated energy landscape. While the success rate of the adiabatic algorithm dramatically drops,  the performance of the hybrid algorithm even improves in the presence of quasi-degeneracies. This is a major benefit of the hybrid scheme compared to adiabatic annealing schemes. It suggests that the algorithm's performance could be enhanced for large samples.

Apart from the final output of the algorithm, other criteria are worth being considered: In contrast to the adiabatic scheme, the hybrid approach does not require long coherence times. 
In view of physical realizations, this favors its scalability. Moreover, since all the relevant data produced in the hybrid algorithm is classical, no quantum-error correction is needed. Since measurements are taken repeatedly, the evolution of the system is permanently tracked. This allows for some feedback control, for instance by automatically stopping the algorithm when the cost function drops below a desired value. Clearly, such strategy produces much more data than a single adiabatic run, providing information also about excited states. This information could be useful if thermodynamic potentials are to be evaluated, and it also may help to evaluate the quality of the final result. 

Moreover, such information could be useful if a large-scale problem is tackled piecewise or in a modular way: Assume a problem which is too big for a single quantum simulator. While there is no obvious way how to perform the adiabatic approach for a problem which exceeds the size permitted by the quantum simulator, the hybrid approach could simply divide the total Hamiltonian into several small clusters, and evolve each cluster separately in a quantum field. Based on the result for each cluster, the classical device can calculate the total cost function, and suggest a new configuration for the full problem.

These advantages compared to an adiabatic quantum annealer may motivate the additional efforts required for the development of hybrid annealer. In particular, for its efficiency it will be crucial to have fast ways of preparing quantum states in the computational basis and of performing measurements projecting the states onto the computational basis.

\acknowledgments{
\textit{Acknowledgements.}
We thank  Alireza Seif, Bruno Juli{\'a}-D{\'i}az, and Stephen Jordan for fruitful discussions. 
We acknowledge financial support from  AFOSR-MURI, from Fundaci\'{o} Cellex, from the European Union
(ERC-2013-AdG Grant No. 339106 OSYRIS, FP7-ICT-2011-9 No. 600645
SIQS, H2020-FETPROACT-2014 No. 641122 QUIC, FP7/2007-2013 Grant
No. 323714 Equam), from Spanish MINECO (FIS2013-46768-P FOQUS, SEV-2015-0522 Severo Ochoa), from the Generalitat de Catalunya
(2014 SGR 874). TG acknowledges a JQI Postdoctoral Fellowship.}


\end{document}